\documentclass[aip,amsmath,amssymb,reprint]{revtex4-1}
\usepackage{graphicx}
\usepackage{dcolumn}
\usepackage{bm}
\usepackage[utf8]{inputenc}
\usepackage[T1]{fontenc}
\usepackage{mathptmx}
\usepackage{etoolbox}
\usepackage[hidelinks]{hyperref} 

\makeatletter
\def\@email#1#2{%
 \endgroup
 \patchcmd{\titleblock@produce}
  {\frontmatter@RRAPformat}
  {\frontmatter@RRAPformat{\produce@RRAP{*#1\href{mailto:#2}{#2}}}\frontmatter@RRAPformat}
  {}{}
}%

\preto\titleblock@produce{%
  \begin{center}
    \small
    This article may be downloaded for personal use only. Any other use requires prior permission of the author and AIP Publishing.\\
    This article appeared in \textit{The Journal of Chemical Physics} \textbf{163}, 084709 (2025) and may be found at
    \href{https://doi.org/10.1063/5.0283532}{https://doi.org/10.1063/5.0283532}.
    \par\vspace{0.8\baselineskip}
  \end{center}%
}
\makeatother
\begin{document}

\preprint{AIP/123-QED}

\title{Sliding properties of Transition Metal Dichalcogenide bilayers}

\author{Pier Luigi Silvestrelli}
\email[]{pierluigi.silvestrelli@unipd.it}
\affiliation{Dipartimento di Fisica e Astronomia, Universit\`a degli Studi di 
Padova, 35131 Padova, Italy}
\author{S. Subashchandrabose}
\affiliation{Centre for Global Health Research, Saveetha Medical College 
\& Hospital, Saveetha Institute of Medical and Technical Sciences (SIMATS),
Chennai-602105, India}
\author{Alberto Ambrosetti}
\affiliation{Dipartimento di Fisica e Astronomia, Universit\`a degli Studi di 
Padova, 35131 Padova, Italy}
\author{Maria Clelia Righi}
\affiliation{Department of Physics and Astronomy, University of Bologna, 
40126 Bologna, Italy}

\newpage
\begin{abstract}
\noindent
Transition-metal dichalcogenides (TMDs) are valuable as solid lubricants 
because of their layered structure, which allows for easy shearing along 
the basal planes. 
Using Density Functional Theory (DFT) we conducted a first-principles study 
of the sliding properties of several TMD bilayers:
MoS$_2$, MoTe$_2$, WS$_2$, WSe$_2$, VS$_2$, VSe$_2$, TaS$_2$, TaSe$_2$, 
TiS$_2$, TiSe$_2$, HfS$_2$, ZrS$_2$, MoS$_2$WS$_2$, MoS$_2$VS$_2$.
Given the crucial role of van der Waals (vdW) interactions in accurately 
describing the interlayer interactions in TMD bilayers, we employed 
vdW-corrected DFT functionals. 
Our research confirms the dominance of vdW effects by estimating the fraction of 
interlayer binding energy attributable to these interactions. 
We also examined how the choice of different vdW-corrected DFT functionals might 
influence quantitative results.
Using MoS$_2$ as a reference TMD bilayer system, we found that most other TMD
bilayers studied exhibit stronger interlayer bonds and greater corrugation. 
However, TiSe$_2$ shows a profile similar to MoS$_2$, while, interestingly,
TiS$_2$, VS$_2$, and ZrS$_2$ are characterized by weaker bonding and 
lower corrugation than MoS$_2$.
We explored relationships between various properties of TMD bilayers, with a 
particular focus on potential connections between tribological and electronic 
properties often characteristic of solid interfaces. To this end, we evaluated 
adhesion energies, work of separation, charge density 
redistributions in interface regions, differential charge densities, 
and corrugation.
While corrugation and thus resistance to sliding generally tends to 
increase with the size of the chalcogen element and is typically proportional 
to the adhesion energy, the relationships between other structural, energetic, 
and electronic properties do not follow a single, well-defined trend.
\end{abstract}

\maketitle
\newpage                                                                             
\section{Introduction}
\noindent
Transition metal dichalcogenides (TMDs) are two-dimensional systems with
unique properties that make them central to the current research in solid-state 
science.\cite{LM}
In TMDs, a transition metal atom (M) layer is sandwiched between two chalcogen 
atom (X) layers, and it is commonly assumed that the MX$_2$ slabs are stacked 
by van der Waals (vdW) interactions, whereas the intralayer 
M-X interactions are covalent. 
The relatively weak, interlayer vdW interactions play a key role in the 
formation, intercalation, exfoliation and layer-by-layer building of TMD 
materials, as well as being decisive for their characteristic sliding properties.

The mechanisms underlying the low friction exhibited
by TMD materials have been investigated both by first-principles
calculations\cite{Cahangirov,Peelaers,Hromadova,Liang,Blumberg,Wang,Levita16,Levita17,Stella} 
and classical molecular dynamics simulations.\cite{Onodera,Stefanov} 
These studies focused on MoS$_2$ because of its excellent performance
for application in devices.
Herein we extend the study to the sliding properties of several TMD bilayers: 
MoS$_2$, MoTe$_2$, WS$_2$, WSe$_2$, VS$_2$, VSe$_2$, TaS$_2$, TaSe$_2$, 
TiS$_2$, TiSe$_2$, HfS$_2$, ZrS$_2$, MoS$_2$WS$_2$, MoS$_2$VS$_2$.
Our research confirms the dominance of vdW effects by estimating the fraction of 
the interlayer binding energy attributable to these interactions. 
We also examined how the choice of different vdW-corrected DFT functionals might 
influence quantitative results.
Using MoS$_2$ as a reference TMD bilayer system, we found that most other TMD
bilayers studied exhibit stronger interlayer bonds and greater corrugation. 
However, TiSe$_2$ shows a profile similar to MoS$_2$, while TiS$_2$, VS$_2$, 
and ZrS$_2$ are characterized by weaker bonding and less corrugation than MoS$_2$.
We explored relationships between various properties of TMD bilayers, with a 
particular focus on potential connections between tribological and electronic 
properties often characteristic of solid interfaces. To this end, we evaluated 
interlayer binding energies (also known as adhesion energies), charge density 
redistributions in interface regions, differential charge densities, and corrugation.

\section{Method}
\noindent
Our first principles study was performed using the Quantum ESPRESSO (QE) 
package,\cite{QE1,QE2} based on the DFT.\cite{dft1,dft2,parr}
In most of the calculations the PBE\cite{PBE} exchange-correlation functional 
was employed, combined with the semiempirical Grimme correction\cite{grimme2} to
properly describe vdW interactions, as done in previous studies on 
TMDs;\cite{Levita14,Levita} hereafter we refer to this functional as
DFT-D2.  
However, other vdW-corrected DFT functionals have been also used (see
Results section) to assess the effect of alternative schemes.
Clearly, the importance of vdW interactions in the characterization of 
layered TMD systems and more generally in materials cannot
be overemphasized (see, for istance, refs. \onlinecite{Ricci,Zhou}).

Wavefunctions were expanded in plane waves with a kinetic energy cutoff 
in the range 50-55 Ry, depending on the constituents atoms.
A $12\times 12\times 12$ and a $12\times 12\times 1$ $k$-point Monkhorst-Pack 
grid was used to sample the Brillouin Zone for TMD bulk and bilayer structures,
respectively.  
In all systems an hexagonal cell was used, containing a MX$_2$ unit per
layer (M=transition metal atom, X=chalcogen atom).
When monolayers or bilayers were modeled, the $c$ dimension
of the cell was enlarged so to allow at least 12 \AA\ of vacuum
between replicated images: the number of $k$ points along such
direction was consequently reduced to one.

The interlayer binding energy $E_b$ is obtained as 
\begin{equation}
E_b = E_{12} - E_1 - E_2\,
\end{equation}
where $E_{12}$ is the energy of the optimized bilayer
and E$_1$ and E$_2$ the energies of the single optimized layers.
The work of separation $W_{\rm sep}$, defined as the energy per unit
area required to separate two layers from equilibrium to infinite
distance, is calculated as $W_{\rm sep} = |E_b|/A$, where $A$ is the unit-cell
area. $W_{\rm sep}$ can therefore be properly compared among systems
having different cell sizes.
The maximum sliding corrugation is indicated as $W_{\rm max}$.

\section{Results and discussions}
\noindent
Here we present the results of our first-principles study 
(mainly based on the DFT-D2 functional\cite{grimme2}) 
of selected TMD bilayers, namely:
MoS$_2$, MoTe$_2$, WS$_2$, WSe$_2$, VS$_2$, VSe$_2$, TaS$_2$, TaSe$_2$, 
TiS$_2$, TiSe$_2$, HfS$_2$, ZrS$_2$, MoS$_2$WS$_2$, MoS$_2$VS$_2$.

The most common configurations assumed by the TMDs are the
{\it 1T-trigonal} and {\it 2H-octahedral} ones (see Fig. 1). 
{\it 1T-trigonal} is
characterized by a trigonal symmetry with a single layer of transition metal 
atoms arranged between two layers of chalcogen atoms in an octahedral 
geometry, while {\it 2H-octahedral} is characterized by an hexagonal
symmetry with two layers of transition metal atoms arranged between layers
of chalcogen atoms in a trigonal prismatic geometry. 
The latter configuration is generally more stable and common for many TMDs,
including (within our list) MoS$_2$, WS$_2$, WSe$_2$, MoTe$_2$; instead
{\it 1T-trigonal} is metastable for many TMDs but it is favored
in a few cases, such as TiS$_2$, TaS$_2$, TaSe$_2$, and HfS$_2$. 
VS$_2$ and VSe$_2$ are reported to prefer the {\it 1T-trigonal} form but 
can also exist in the {\it 2H-octahedral} one under certain conditions. 
These preferences are mainly due to the electronic configuration of the 
transition metal atoms and their interactions with the chalcogen atoms. 
Differently from, for instance, graphene, where two commensurate planes
always show the same relative orientation, TMDs have an
additional degree of freedom when forming the bilayer: in fact the
layers can be either {\it parallel} or {\it antiparallel}. Here we
have investigated both orientations, labeled R0
and R180, respectively.\cite{Levita14,Levita}
In fact, while most investigations
on TMDs have dealt with the R180 structure (typical of
the 2H polytype), examples of 3R polytypes presenting the
R0 orientation were found in both natural and synthetic 
crystals.\cite{He,Winer} A previous study\cite{Levita14} showed
that for MoS$_2$ R0 and R180 bilayers are degenerate at
zero applied load while the degeneracy is instead removed by
uniaxial stress.

\begin{figure}
\renewcommand{\figurename}{FIG.}
\renewcommand{\thefigure}{\arabic{figure}(a)}
\centerline{
\includegraphics[width=8cm,angle=270]{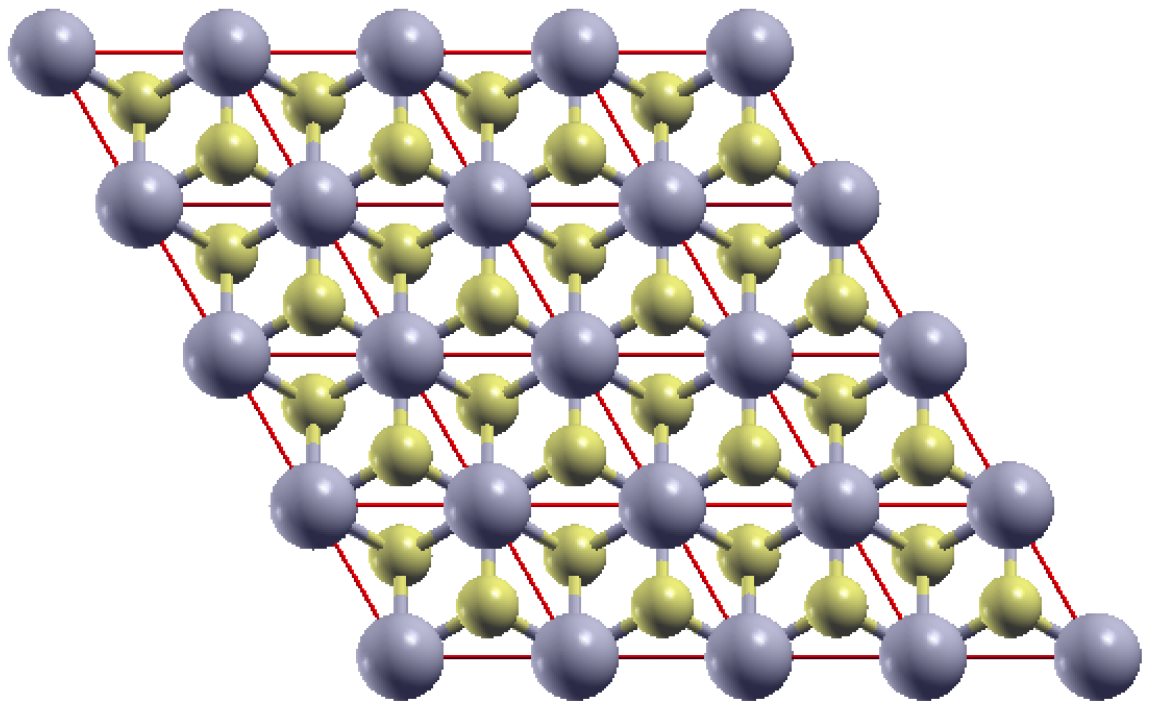}
}
{\vskip -2.8cm}
\includegraphics[width=8cm,angle=0]{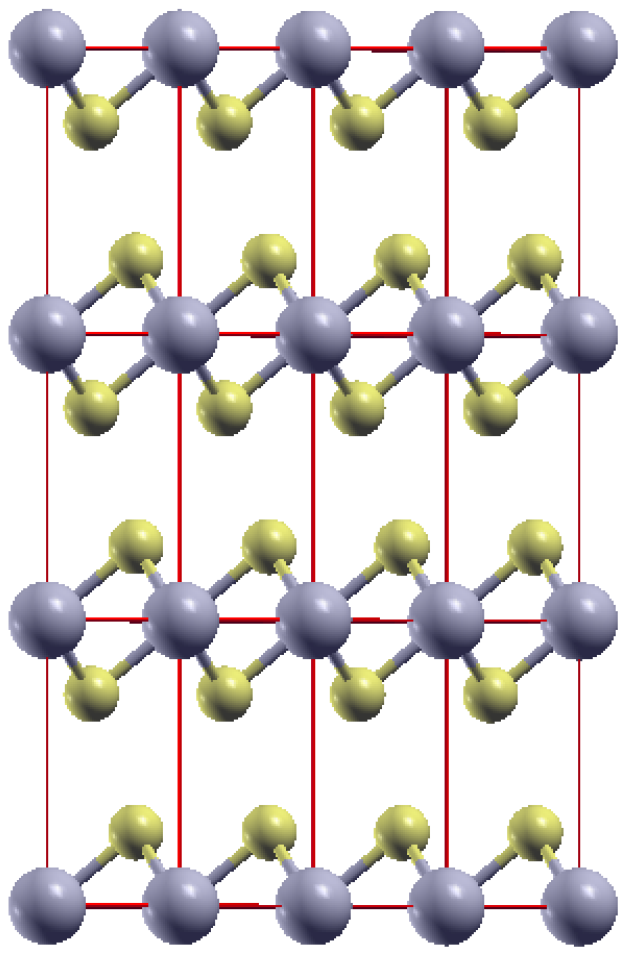}
{\vskip -2.2cm}
\caption{Bulk MoS$_2$ in {\it 1T-trigonal} configuration.
Upper figure: top view; lower figure: side view.
The unit cells are marked with red solid lines; Mo atoms are
represented in gray and S atoms in yellow.}
\label{fig1a}
\end{figure}

\begin{figure}
\renewcommand{\figurename}{FIG.}
\renewcommand{\thefigure}{1(b)}
\centerline{
\includegraphics[width=8cm,angle=270]{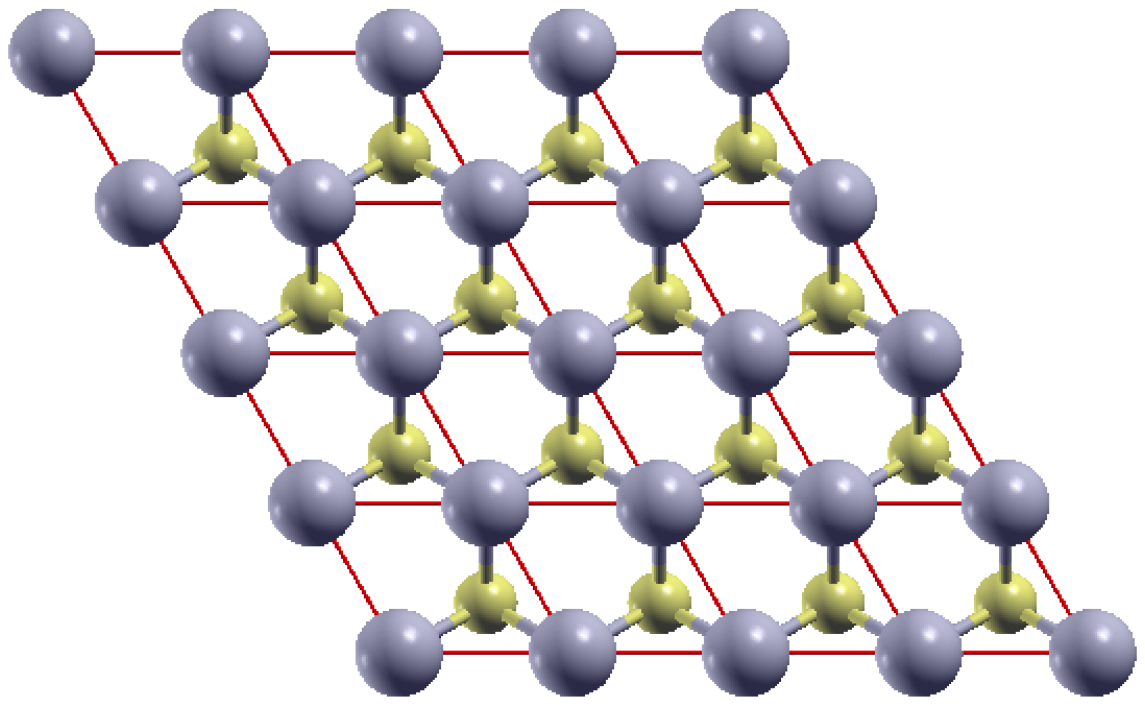}
}
{\vskip -2.0cm}
\includegraphics[width=7cm,angle=0]{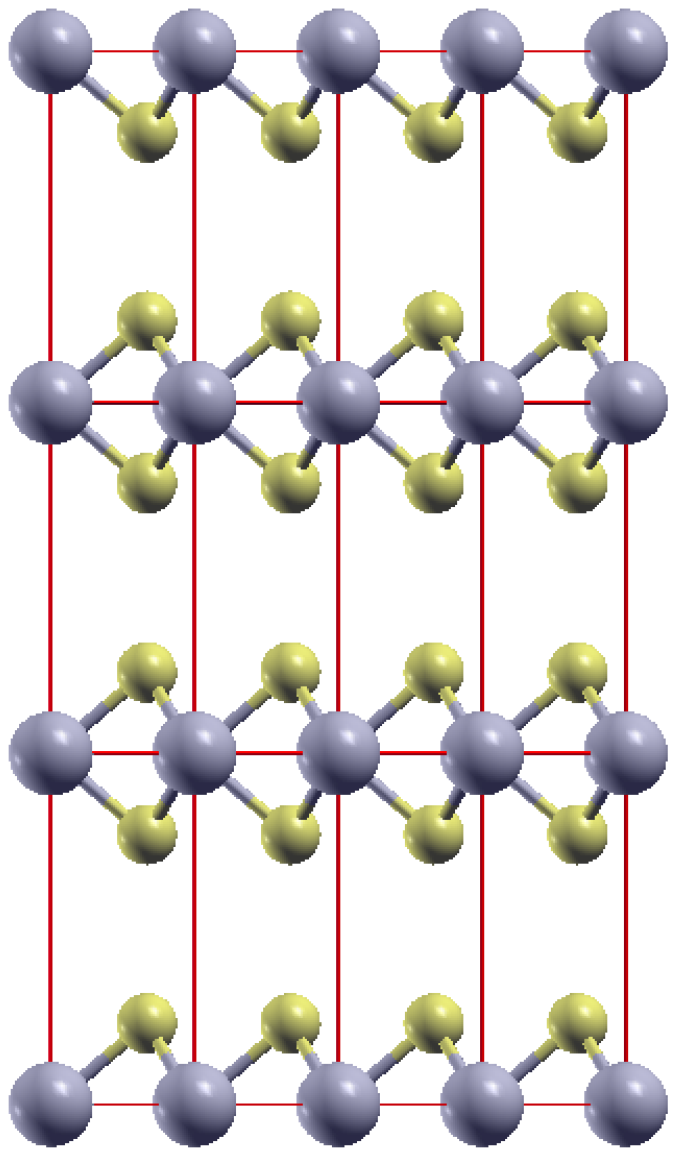}
{\vskip -1.5cm}
\caption{Bulk MoS$_2$ in {\it 2H-octahedral} configuration.
Upper figure: top view; lower figure: side view.
The unit cells are marked with red solid lines; Mo atoms are
represented in gray and S atoms in yellow.}
\label{fig1b}
\end{figure}

\begin{table}
\caption{Calculated values for the lattice constant $a$, interlayer
distance $d_{\rm int}$, work of separation $W_{\rm sep}$, and maximum 
sliding corrugation $W_{\rm max}$ for TMD bilayers in {\it 2H-octahedral}
configuration. Distances are in \AA\ and energies
in J/m$^2$; R0 or R180 denote parallel or antiparallel
orientation of bilayers. 
Where available, experimental (from refs. \onlinecite{Komsa}) 
and RPA reference values 
(from ref. \onlinecite{He}) are reported in parentheses for comparison.}
\begin{ruledtabular}
\begin{tabular}{|c|c|c|c|c|}
       system  & $a$   & $d_{\rm int}$  & $W_{\rm sep}$ & $W_{\rm max}$ \\ 
\hline
 MoS$_2$  R0   & 3.198 & 6.208 & 0.277 & 0.100 \\
 MoS$_2$  R180 & 3.198 & 6.218 & 0.277 & 0.096 \\
               &(3.160)& (6.27)& (0.300) & (0.127) \\
\hline
 MoTe$_2$ R0   & 3.524 & 7.036 & 0.484 & 0.202 \\
 MoTe$_2$ R180 & 3.524 & 7.031 & 0.505 & 0.215 \\
\hline
 WS$_2$   R0   & 3.180 & 6.006 & 0.497 & 0.167 \\
 WS$_2$   R180 & 3.180 & 6.179 & 0.497 & 0.161 \\
               &(3.153)& (6.24)& (0.308) & (0.141) \\
\hline
 WSe$_2$  R0   & 3.327 & 6.402 & 0.569 & 0.200 \\
 WSe$_2$  R180 & 3.327 & 6.466 & 0.575 & 0.198 \\
               &(3.282)& (6.50)& (0.308) & (0.136) \\
\hline
 VS$_2$   R0   & 3.175 & 6.045 & 0.247 & 0.091 \\
 VS$_2$   R180 & 3.175 & 6.078 & 0.252 & 0.095 \\
\hline
 VSe$_2$  R0   & 3.328 & 6.383 & 0.327 & 0.120 \\
 VSe$_2$  R180 & 3.328 & 6.410 & 0.324 & 0.114 \\
\hline
 TaS$_2$  R0   & 3.345 & 6.002 & 0.486 & 0.151 \\
 TaS$_2$  R180 & 3.345 & 5.915 & 0.503 & 0.167 \\
\hline
 TaSe$_2$ R0   & 3.477 & 6.330 & 0.529 & 0.185 \\
 TaSe$_2$ R180 & 3.477 & 6.256 & 0.542 & 0.198 \\
\hline
 TiS$_2$  R0   & 3.329 & 6.145 & 0.201 & 0.065 \\
 TiS$_2$  R180 & 3.329 & 5.954 & 0.177 & 0.091 \\
\end{tabular}
\end{ruledtabular}
\label{table1}
\end{table}

\begin{table}
\vfill
\eject
\caption{Calculated values for the lattice constant $a$, interlayer
distance $d_{\rm int}$, work of separation $W_{\rm sep}$, and maximum 
sliding corrugation $W_{\rm max}$ for TMD bilayers in {\it 1T-trigonal}
configuration. Distances are in \AA\ and energies
in J/m$^2$; R0 or R180 denote parallel or antiparallel
orientation of bilayers.} 
\begin{ruledtabular}
\begin{tabular}{|c|c|c|c|c|}
        system  & $a$   & $d_{\rm int}$  & $W_{\rm sep}$ & $W_{\rm max}$ \\ 
\hline
 MoS$_2$  R0   & 3.227 & 6.009 & 0.316 & 0.123 \\
 MoS$_2$  R180 & 3.227 & 5.990 & 0.334 & 0.142 \\
\hline
 VS$_2$   R0   & 3.193 & 5.947 & 0.258 & 0.093 \\
 VS$_2$   R180 & 3.193 & 5.994 & 0.250 & 0.088 \\
\hline
 VSe$_2$  R0   & 3.344 & 6.288 & 0.341 & 0.123 \\
 VSe$_2$  R180 & 3.344 & 6.319 & 0.340 & 0.125 \\
\hline
 TaS$_2$  R0   & 3.393 & 5.803 & 0.466 & 0.159 \\
 TaS$_2$  R180 & 3.393 & 5.886 & 0.455 & 0.151 \\
\hline
 TaSe$_2$ R0   & 3.509 & 6.174 & 0.550 & 0.193 \\
 TaSe$_2$ R180 & 3.509 & 6.250 & 0.534 & 0.179 \\
\hline
 TiS$_2$  R0   & 3.392 & 5.852 & 0.218 & 0.078 \\
 TiS$_2$  R180 & 3.392 & 5.952 & 0.196 & 0.055 \\
\hline
 TiSe$_2$ R0   & 3.510 & 6.188 & 0.315 & 0.120 \\
 TiSe$_2$ R180 & 3.510 & 6.277 & 0.288 & 0.091 \\
\hline
 HfS$_2$  R0   & 3.623 & 5.679 & 0.421 & 0.147 \\
 HfS$_2$  R180 & 3.623 & 5.729 & 0.401 & 0.126 \\
\hline
 ZrS$_2$  R0   & 3.632 & 5.935 & 0.200 & 0.069 \\
 ZrS$_2$  R180 & 3.632 & 6.005 & 0.185 & 0.053 \\
\end{tabular}
\end{ruledtabular}
\label{table2}
\end{table}

Tables I and II report the value of the lattice constant $a$ and of
the optimized interlayer distance $d_{\rm int}$ (between 
transition metal atoms M in different layers).
$a$ has been determined by optimizing the cell parameters of
the TMD systems by considering their bulk structure.  
Note that, while our calculations confirm that for bulk MoS$_2$
the {\it 2H-octahedral} structure is energetically favored over
the {\it 1T-trigonal} one (by about 0.7 eV per unit cell), in the case of 
MoS$_2$ bilayer the opposite is found (see Tables I and II).  
For most of the investigated TMD bilayers we have considered both the 
{\it 1T-trigonal} and {\it 2H-octahedral} configurations since the binding
energies of both the structures are comparable.

As can be seen in Tables I and II, compounds containing sulfur (S) tend to have 
lower $a$ values compared to their selenium (Se) or tellurium (Te) analogs,
while compounds with heavier metals than Mo, such as W, Ta, and Hf, tend to have 
higher $a$ values. 
A general, approximate positive correlation exists between $a$ and 
$d_{\rm int}$: compounds with larger $a$ tend to have larger interlayer
distance. 
Moreover, $a$, $d_{\rm int}$, and $W_{\rm sep}$ values increase, 
for the same M element, going from S to Se and to Te: in fact a larger 
polarizability of the interfacial chalcogen atoms leads to stronger
vdW interactions between the layers, in line with previous 
findings.\cite{Levita} Considering the lowest-energy configuration,
WSe$_2$ has the highest $W_{\rm sep}$ value, followed by
TaSe$_2$ and MoTe$_2$, while ZrS$_2$ and TiS$_2$ have the lowest values.
Clearly these correlations can be explained by considering the atomic sizes, 
electronegativity, and electronic structure of the various elements involved.

Interestingly, there is a strong positive correlation between $W_{\rm sep}$ 
and the maximum sliding corrugation $W_{\rm max}$: TMD bilayers with
higher $W_{\rm sep}$ values tend to have higher corrugations:
$W_{\rm max} \sim 0.35 W_{\rm sep}$, which agrees with the general 
understanding that stronger interlayer bonding leads to higher corrugation 
and friction.
Note that a similar correlation was also found\cite{Wolloch} for 
metal interfaces (the cleavage strength is directly proportional to the
adhesion); moreover an expression for the corrugation          
related to the repulsive part of the intralayer adhesion was           
suggested\cite{Reguzzoni} for the graphene bilayer.
Moderate deviations from this trend are represented by VSe$_2$ (smaller 
corrugation than expected) and MoTe$_2$ (larger corrugation than expected).
Tables I and II therefore exhibit several notable correlations among the 
reported TMD properties, with $W_{\rm sep}$ appearing to be a central 
parameter that correlates well with both interlayer distance and 
energy corrugation values, as also observed in previous DFT 
studies.\cite{Irving} 
The choice of metal and chalcogen atoms significantly influences these 
properties as well, with heavier elements generally leading to stronger 
interlayer interactions and more corrugated energy landscapes.
Larger interfacial atoms on the one hand increase the adhesion,
due to the stronger van der Waals interlayer forces; on the
other, they increase the corrugation, due to the Pauli repulsion
between electronic clouds of on-top chalcogen-chalcogen
structures (corresponding to the potential-energy maxima). It was shown that vdW
interactions do not contribute significantly to the
corrugation;\cite{Reguzzoni} in the case of graphene, in fact, a much
lower corrugation with respect to MoS$_2$ is observed, despite
their almost equal adhesion energies.
For almost all systems considered, the corrugation $W_{\rm max}$
is larger than in the case of graphene where $W_{\rm max}=0.060$
J/m$^2$;\cite{Levita} notable exceptions are represented by
TiS$_2$ and ZrS$_2$ bilayers in the {\it 1T-trigonal} configuration
with R180 orientation, which are characterized by a corrugation 
even slightly lower than that of graphene (see Table II).

Note that VS$_2$ and VSe$_2$ have ferromagnetic properties,
as confirmed by our DFT calculations: in fact for VS$_2$
we find a total magnetic moment of about 1.1 and 1.8 Bohr magneton per unit cell
in the {\it 1T-trigonal} and {\it 2H-octahedral} configuration,
respectively, while, for VSe$_2$, the corresponding values are 
1.3 and 1.9 Bohr magneton per unit cell.
Actually these systems are classified as ``strongly correlated materials'',
with VSe$_2$ which is attractive for van der Waals
spintronics applications\cite{Ma,Samad,Bonilla} and 
VS$_2$ which is not very stable;\cite{Samad,Li} 
there are several factors contributing to VS$_2$ instability:
in particular, VS$_2$ monolayers tend to undergo a structural distortion
(structural and dyamical instability); moreover
partially filled {\it d}-orbitals of vanadium
make it prone to electronic instabilities leading to magnetic ordering,
and make VS$_2$ monolayers chemically very active.
Calculations were performed for these system also using a 
more suitable Hubbard LDA+U approach (with U=1.0 eV);\cite{Esters} 
however no significant changes were found with respect to the standard method. 
For instance, in the case of the {\it 2H-octahedral} R0 
configuration of VS$_2$, $W_{\rm max}$ is slightly smaller with 
Hubbard LDA+U, the reduction being of about 9\%.

In Figs. 2 and 3 we plot the variation of the interlayer binding energy
per unit-cell area $E_b/A$ in our TMD bilayers.
Each profile is associated with the sliding of the bilayer, along 
a specified path, as exemplified by 
the $y$ direction indicated by the blue arrow in Fig. 4.
Symbols correspond to calculated data; lines are just guides for the eyes.
Since the R180 structure appears to be preferred by most of
systems in the {\it 2H-octahedral} configuration (which is
typical of the 2H polytype\cite{Levita}), while the R0 
structure is instead favored by most of the systems in
the {\it 1T-trigonal} configuration, we only plot data 
relative to the R180 structure for {\it 2H-octahedral} configurations
(Fig. 2) and data relative to the R0 structure for {\it 1T-trigonal} 
configurations (Fig. 3).
In any case, for a given system, Tables I and II show that the energy 
differences between R0 and R180 structures are at most moderate; 
for instance, we find that in the case of MoS$_2$ R0 and R180 bilayers are
essentially degenerate, in line with previous reports at
zero load.\cite{Levita}
More significant differences between R0 and R180 structures
are instead expected in simulations with an applied load.\cite{Levita}

\begin{figure}
\addtocounter{figure}{-1}
\includegraphics[width=8cm,angle=270]{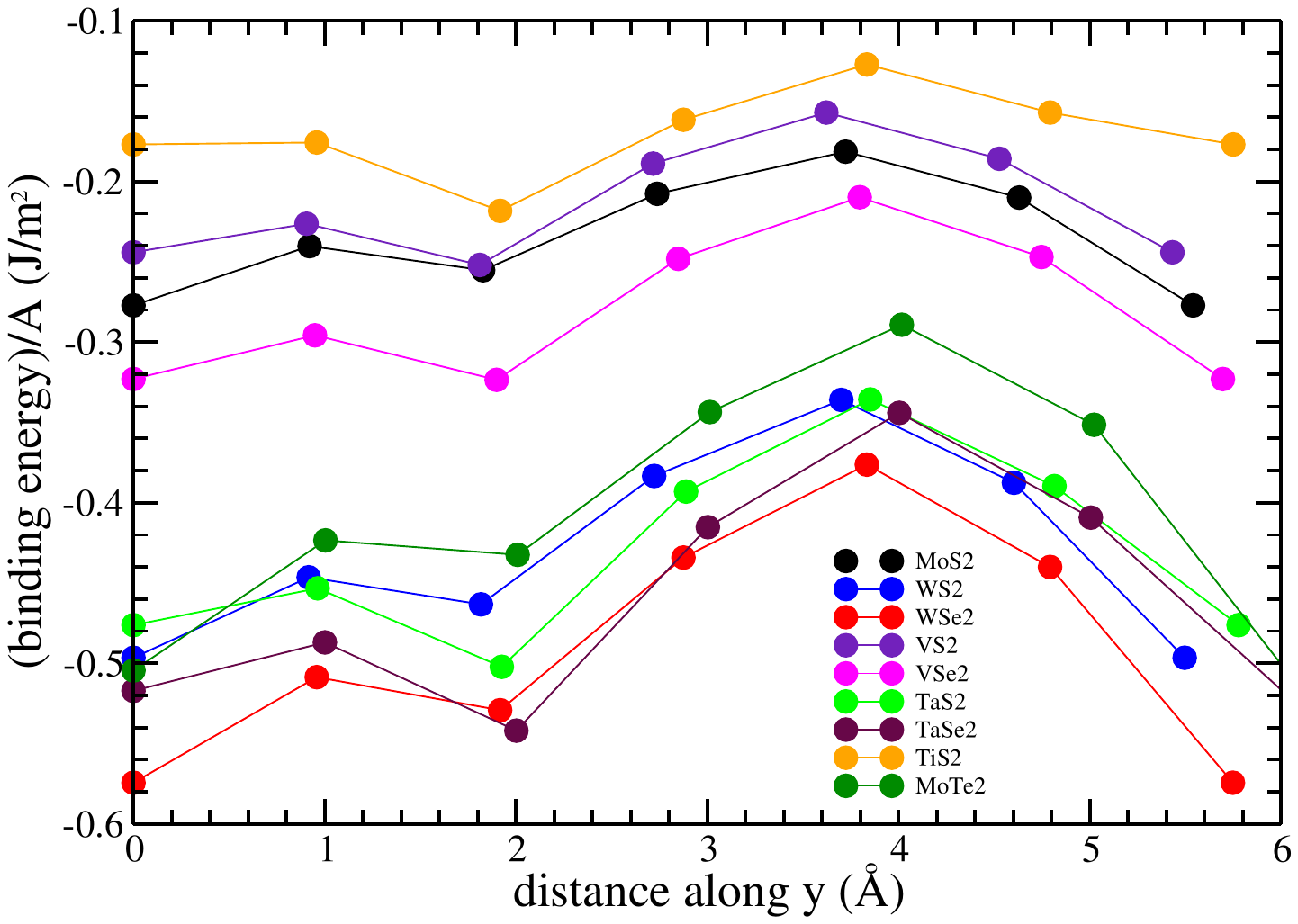}
\caption{Variation of the interlayer binding energy
per unit-cell area, $E_b/A$, in different
TMD bilayers, considering the R180 structure for {\it 2H-octahedral} 
configurations. Each profile is associated with the sliding of the 
bilayer, along the path in the $y$ direction indicated by the blue arrow 
in Fig. 4. 
Symbols correspond to calculated data; lines are just guides for the eyes.}
\label{fig2}
\end{figure}

\begin{figure}
\includegraphics[width=8cm,angle=270]{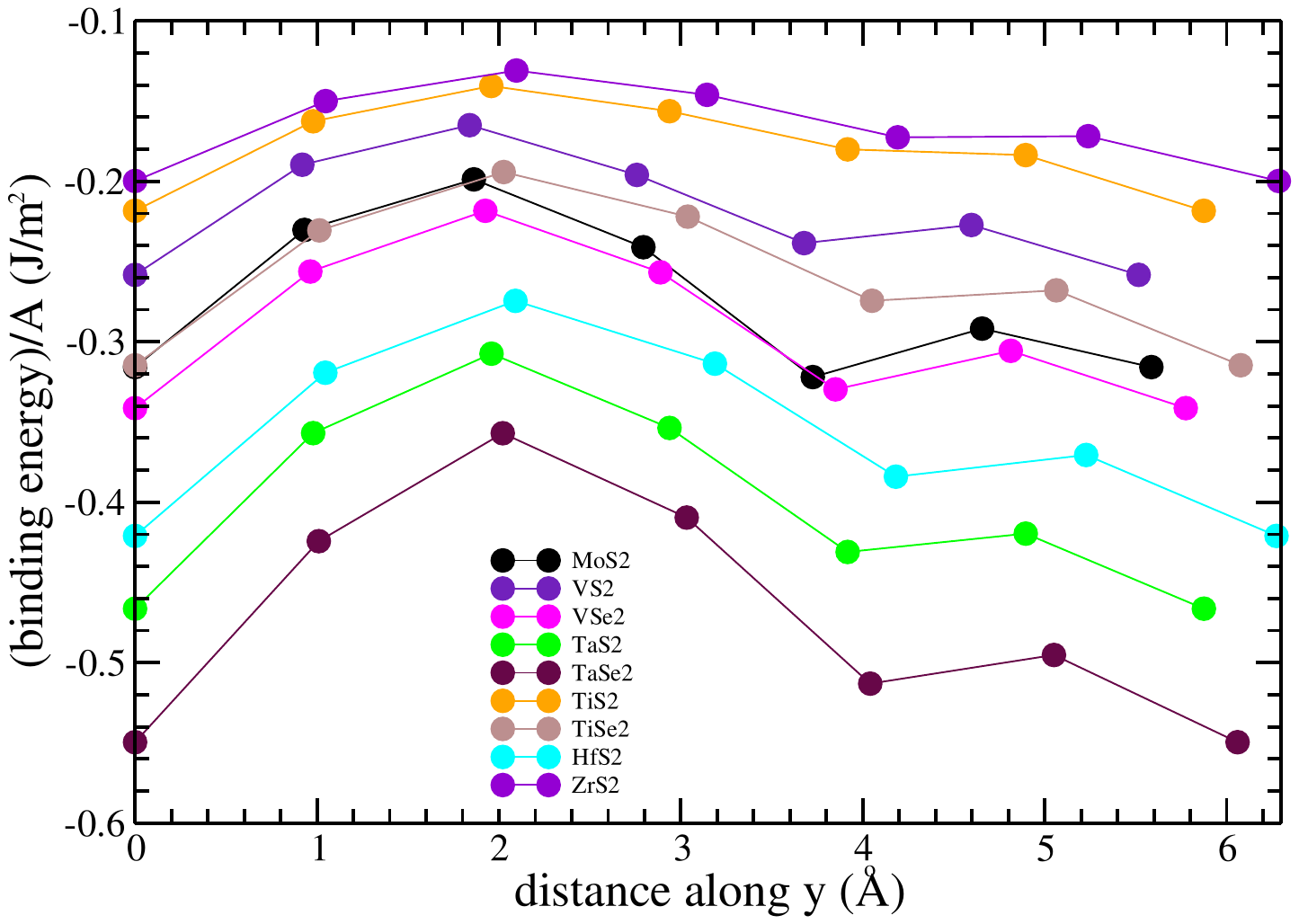}
\caption{Variation of the interlayer binding energy
per unit-cell area, $E_b/A$, in different
TMD bilayers, considering the R0 structure for {\it 1T-trigonal}
configurations. Each profile is associated with the sliding of the 
bilayer, along the path in the $y$ direction indicated by the blue arrow 
in Fig. 4. 
Symbols correspond to calculated data; lines are just guides for the eyes.}
\label{fig3}
\end{figure}

\begin{figure}
\includegraphics[width=8cm,angle=270]{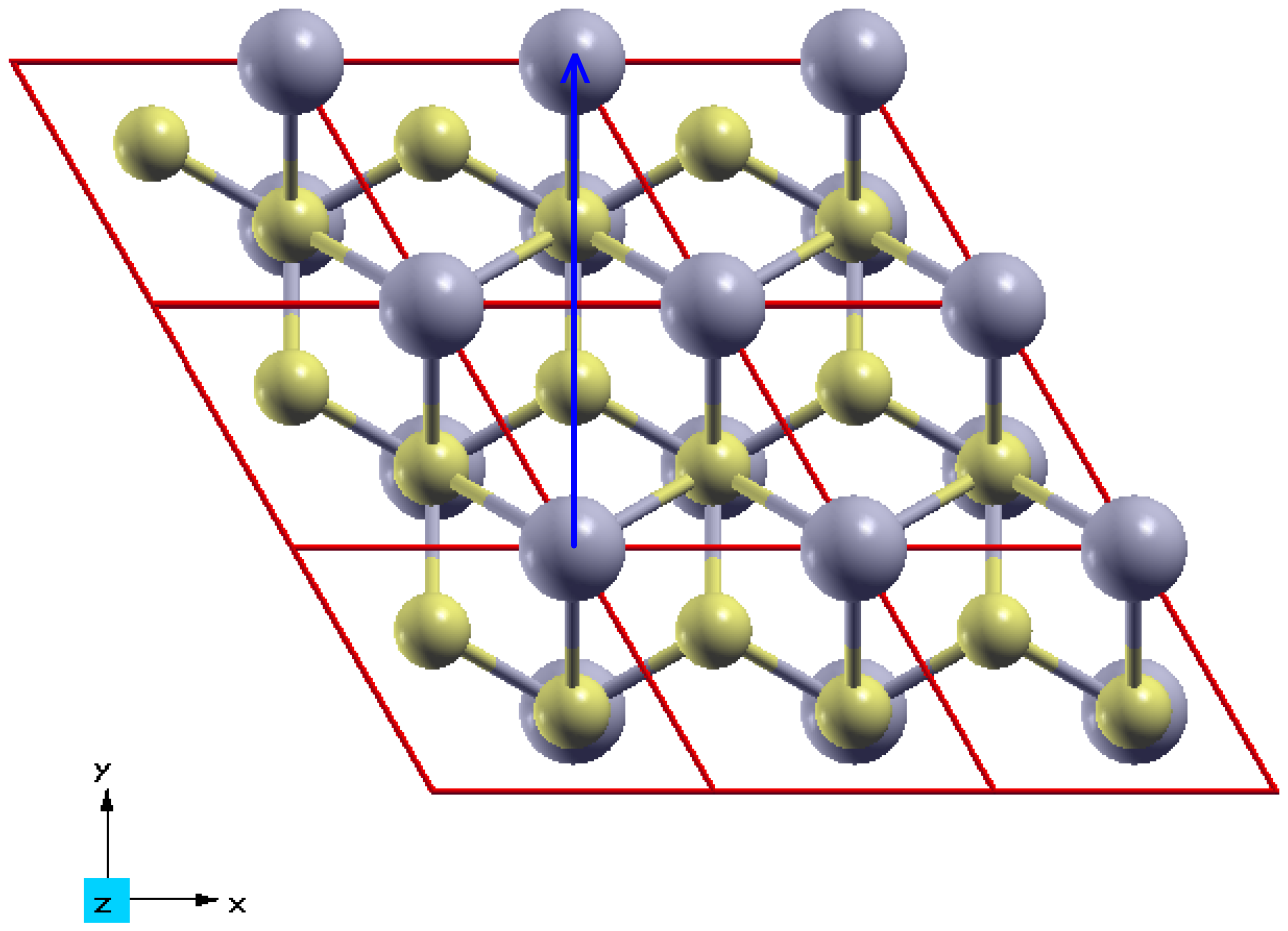}
\caption{Sliding path (indicated by the blue arrow) relative to 
the R0 structure in the {\it 2H-octahedral} configuration of 
MoS$_2$ (top view). 
The unit cells are marked with red solid lines; Mo atoms are
represented in gray and S atoms in yellow.}
\label{fig4}
\end{figure}

Obviously, the more negative the values, the stronger the interaction between 
the two layers.
As can be seen, Figs. 2 and 3 show that (see also Tables I and II)
with respect to the MoS$_2$ case, taken as a reference,
almost all the other bilayers are characterized by stronger interlayer 
bonds and greater corrugation, but for the TiSe$_2$ system which exhibits 
a profile similar to that of MoS$_2$, and for TiS$_2$, VS$_2$, and ZrS$_2$ 
bilayers which appear less bonded and less corrugated than MoS$_2$.  
Interestingly, the TMD bilayers which are less corrugated than MoS$_2$
(TiS$_2$, VS$_2$, ZrS$_2$)
are all characterized by a difference between the Pauling electronegativity
of the chalcogen and the metal element that is larger than in MoS$_2$,
thus partially supporting the conjecture of Irving {\it et al.}\cite{Irving} 
who proposed that intrinsic frictional properties
of TMD compounds can be effectively guessed from a
property as simple as atomic electronegativity: slippage between planes
should be facilitated by the presence of enhanced (slightly decreased) charge
accumulation on the chalcogens (metals).
However, our results suggest that this conclusion is not 
valid in general, as demonstrated by the fact that several TMD bilayers
(HfS$_2$, TaS$_2$, TaSe$_2$, TiSe$_2$, VS$_2$, VSe$_2$)       
exhibit higher (or at most similar) corrugation than MoS$_2$, despite 
a larger difference between the electronegativity of the chalcogen and 
the metal element.
In line with previous DFT studies,\cite{Irving} we find that
larger chalcogen atoms lead to stronger interlayer binding.
Also note that, in the case of the R180 structure for {\it 2H-octahedral}
configurations, while in most systems we considered, the starting position
coincides with the lowest-energy minimum (characterized by the chalcogen
atoms in the top layer sitting directly above the metal atoms in the bottom 
layer\cite{Levita,Irving}), in the case of TaS$_2$, TaSe$_2$,
VS$_2$, and VSe$_2$ the structure corresponding to a displaced position
is instead energetically favored.

\begin{table}
\vfill
\eject
\caption{Calculated values for the lattice constant $a$, interlayer
distance $d_{\rm int}$, work of separation $W_{\rm sep}$, and maximum 
sliding corrugation $W_{\rm max}$ considering the {\it 2H-octahedral}
configuration of TaS$_2$ with R0 or R180 orientation of bilayers, using
different vdW-corrected DFT functionals. Distances are in \AA\ and energies
in J/m$^2$.}
\begin{center}
\begin{tabular}{|c|c|c|c|c|}
\hline
   DFT functional  & $a$   & $d_{\rm int}$  & $W_{\rm sep}$ & $W_{\rm max}$ \\ 
\hline
 DFT-D2  R0   & 3.345 & 6.002 & 0.486 & 0.151 \\
 DFT-D2  R180 & 3.345 & 5.915 & 0.503 & 0.167 \\
\hline
 DFT-D3  R0   & 3.319 & 6.163 & 0.348 & 0.094 \\
 DFT-D3  R180 & 3.319 & 6.193 & 0.355 & 0.101 \\
\hline
 rVV10   R0   & 3.351 & 6.147 & 0.442 & 0.120 \\
 rVV10   R180 & 3.351 & 6.194 & 0.454 & 0.131 \\
\hline
\end{tabular}
\end{center}
\label{table3}
\end{table}

In order to assess to what extent our reported quantitative results 
can depend on the choice of the adopted DFT functional, 
in Fig. 5 and Table III we compare, in the selected 
case of the TaS$_2$ TMD bilayer (considering the {\it 2H-octahedral}
configuration), the energy profiles obtained by DFT-D2 to those
obtained by other vdW-corrected DFT functionals: DFT-D3,\cite{DFT-D3}
in which the molecular environment of a given atom is taken into account
by the empirical concept of fractional coordination number, and the 
rVV10 functional\cite{rVV10} (the revised, more efficient
version of the original VV10 scheme\cite{VV10}), where
vdW effects are included by introducing an explicitly nonlocal correlation
functional.
As can be seen, the energy profiles of all the three functionals 
exhibit a similar qualitative behavior; however significant 
quantitative differences are evident.
The DFT-D3 binding energies are quite smaller (in absolute value) than
the DFT-D2 ones, while instead rVV10 gives intermediate values 
between DFT-D3 and DFT-D2, being however closer to those of DFT-D2.  
By comparing DFT-D3 with DFT-D2, from Table III, one can also see that 
with DFT-D3 the weaker bonding is consistent with
larger interlayer distances, while instead, with the 
rVV10 functional, the interlayer distances are comparable with those
of DFT-D3.     
Similar changes in binding energy, using the three different DFT
functionals, were also observed in the TiS$_2$ bilayer. 
Therefore the choice of a specific vdW-corrected DFT functional
can significantly influence the quantitative results, as already 
discussed in other theoretical studies on TMDs (see, for instance,
ref. \onlinecite{Bjorkman}).

\begin{figure}
\centerline{
\includegraphics[width=8cm,angle=270]{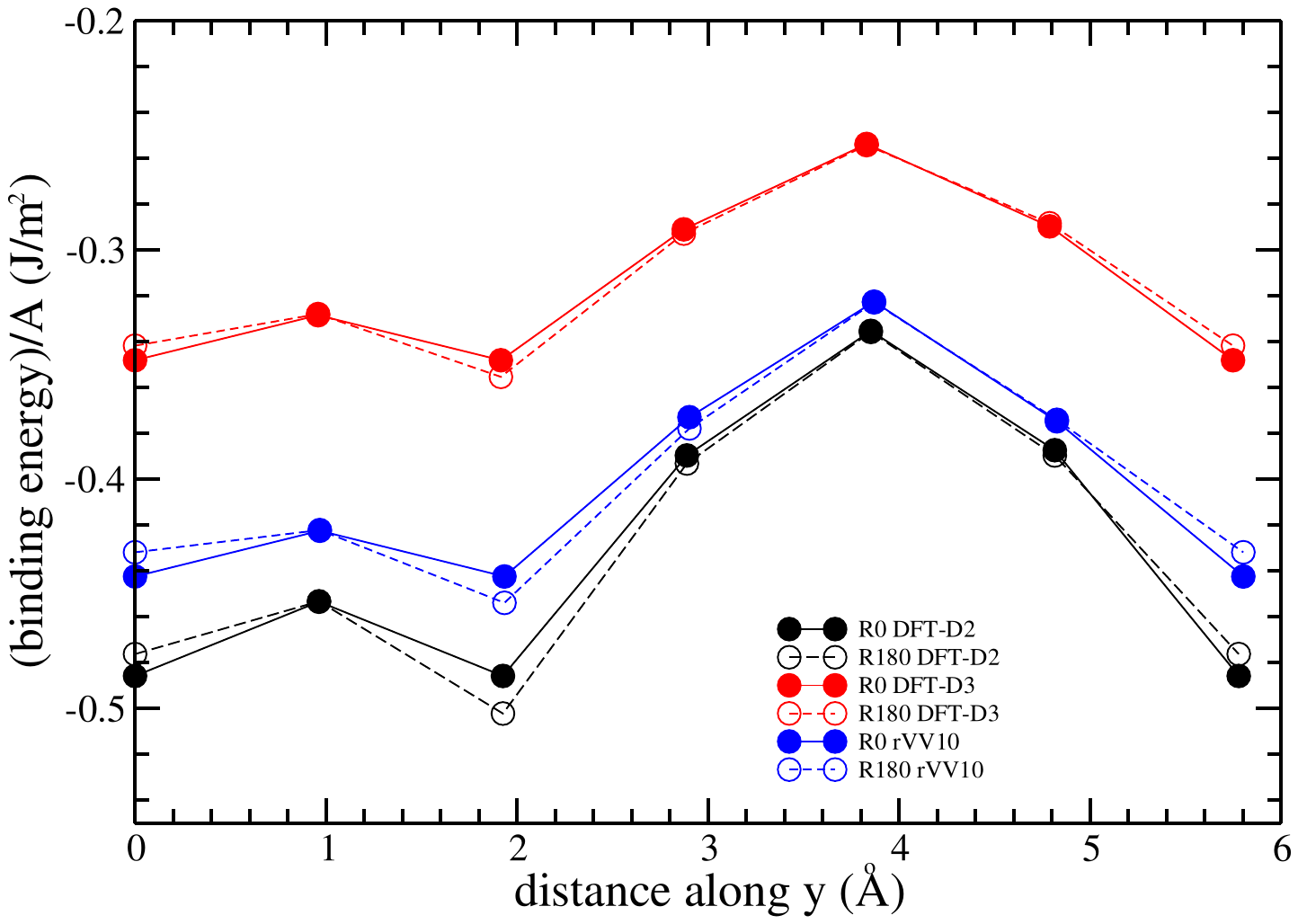}
}
\caption{Variation of the interlayer binding energy
per unit-cell area, $E_b/A$, relative to R0 and R180 structures of the
TaS$_2$ bilayer (considering the {\it 2H-octahedral}
configuration), obtained by different vdW-corrected DFT functionals:
DFT-D2, DFT-D3, and rVV10 (see text).
Symbols correspond to calculated data; lines are just guides for the eyes.}
\label{fig5}
\end{figure}

By using the semiempirical DFT-D2 and DFT-D3 functionals, one 
can also easily estimate the fraction of the interlayer binding energy that 
can be ascribed to vdW interactions. 
As expected for this kind of bilayer systems, the
vdW interactions are predominant (see Fig. 6, again for the TaS$_2$ system):
quantitatively, at the optimal, equilibrium interlayer distance,
the vdW fraction of the binding energy is 98\% and 110\% with DFT-D2 
and DFT-D3, respectively; in the latter case this means that the
energy due to other types of interactions is positive (repulsive contribution).

\begin{figure}
\includegraphics[width=8cm,angle=270]{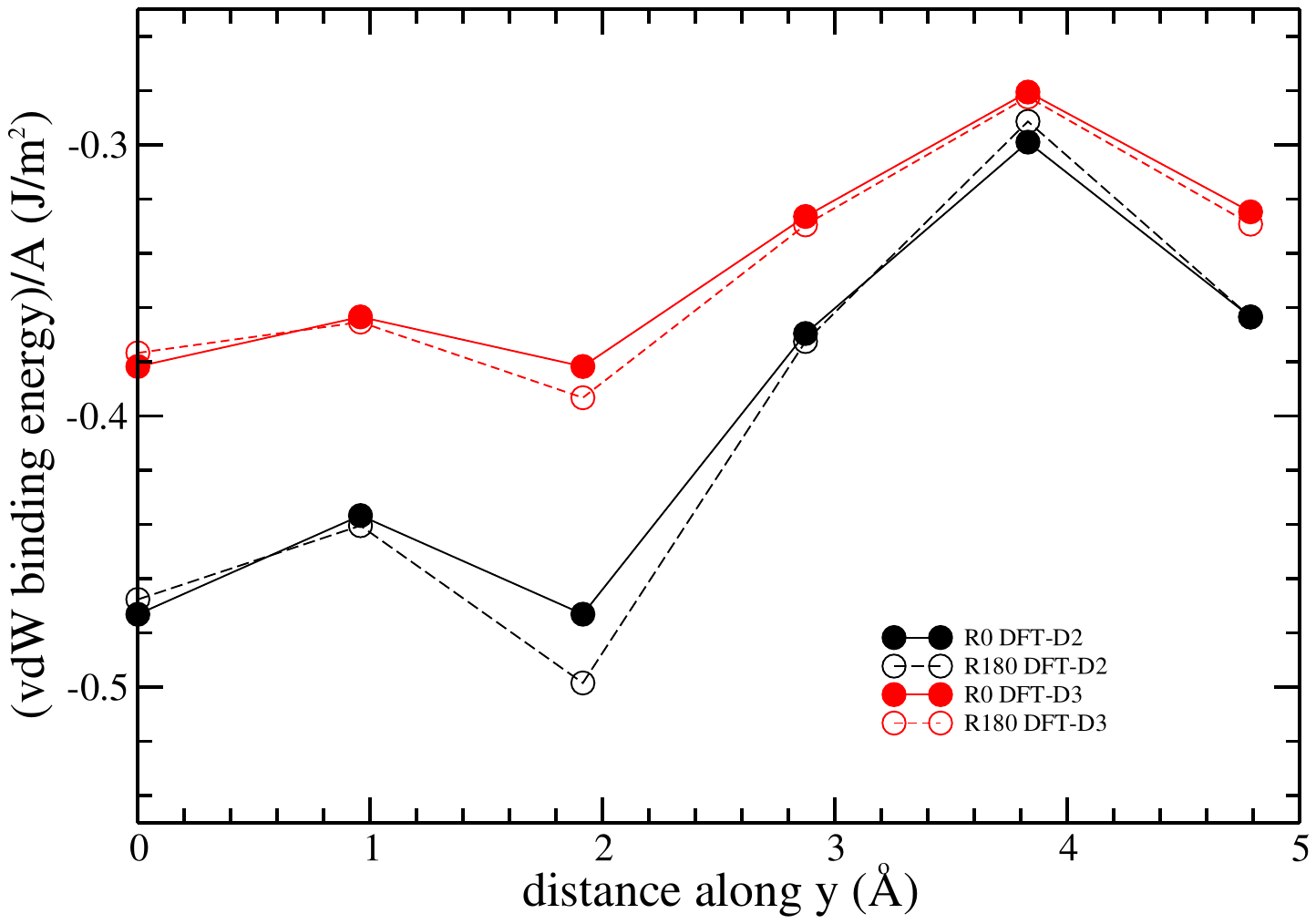}
\caption{Variation of the interlayer binding energy
per unit-cell area, considering the vdW contribution only, 
relative to R0 and R180 structures of the
TaS$_2$ bilayer (in the {\it 2H-octahedral} configuration), obtained by 
the DFT-D2 and DFT-D3 functionals.
Symbols correspond to calculated data; lines are just guides for the eyes.}
\label{fig6}
\end{figure}

Interestingly, in ref.\onlinecite{Righi-prl-2018} a connection between the 
intrinsic tribological properties and the electronic properties of a
solid interface was pointed out. In particular, the adhesion and frictional 
forces seem to be determined by the electronic charge redistribution 
occurring due to the relative displacements of the two surfaces in contact.
To verify to what extent this conclusion is applicable to our
investigated systems, the relations between different quantities
have been considered for selected TMD bilayers with the R180 structure
in the {\it 2H-octahedral} configuration.
In Figs. S1 and S2 in supplementary material, and in Fig. 7 
the work of separation $W_{\rm sep}$ is
related to the redistribution of charge density in the interface region
($\rho_{\rm redist}$, as defined in ref.\onlinecite{Righi-prl-2018}), 
the height $h$ of the central peak of the planar average of the differential
density\cite{Righi-prl-2018} and the
maximum sliding corrugation $W_{\rm max}$, respectively.
Analysis of these figures indicates that:
{\it i)} $W_{\rm max}$ turns out to be proportional to $W_{\rm sep}$
(Fig. 7), in line with the general behavior observed above and with
the results of ref.\onlinecite{Righi-prl-2018};
{\it ii)} $\rho_{\rm redist}$ and $h$ are instead proportional to 
$W_{\rm sep}$ only considering a given M element (for instance Ta in 
TaS$_2$ and TaSe$_2$, and W in WS$_2$ and WSe$_2$) and changing the chalcogen 
element, while the same is not true considering a given chalcogen element 
(for instance S in MoS$_2$, WS$_2$ and TaS$_2$)    
and changing the M element; note however the exception represented by
the VS$_2$ and VSe$_2$ systems, since $\rho_{\rm redist}$ is larger for 
VS$_2$ than for VSe$_2$.

\begin{figure}
\includegraphics[width=8cm,angle=270]{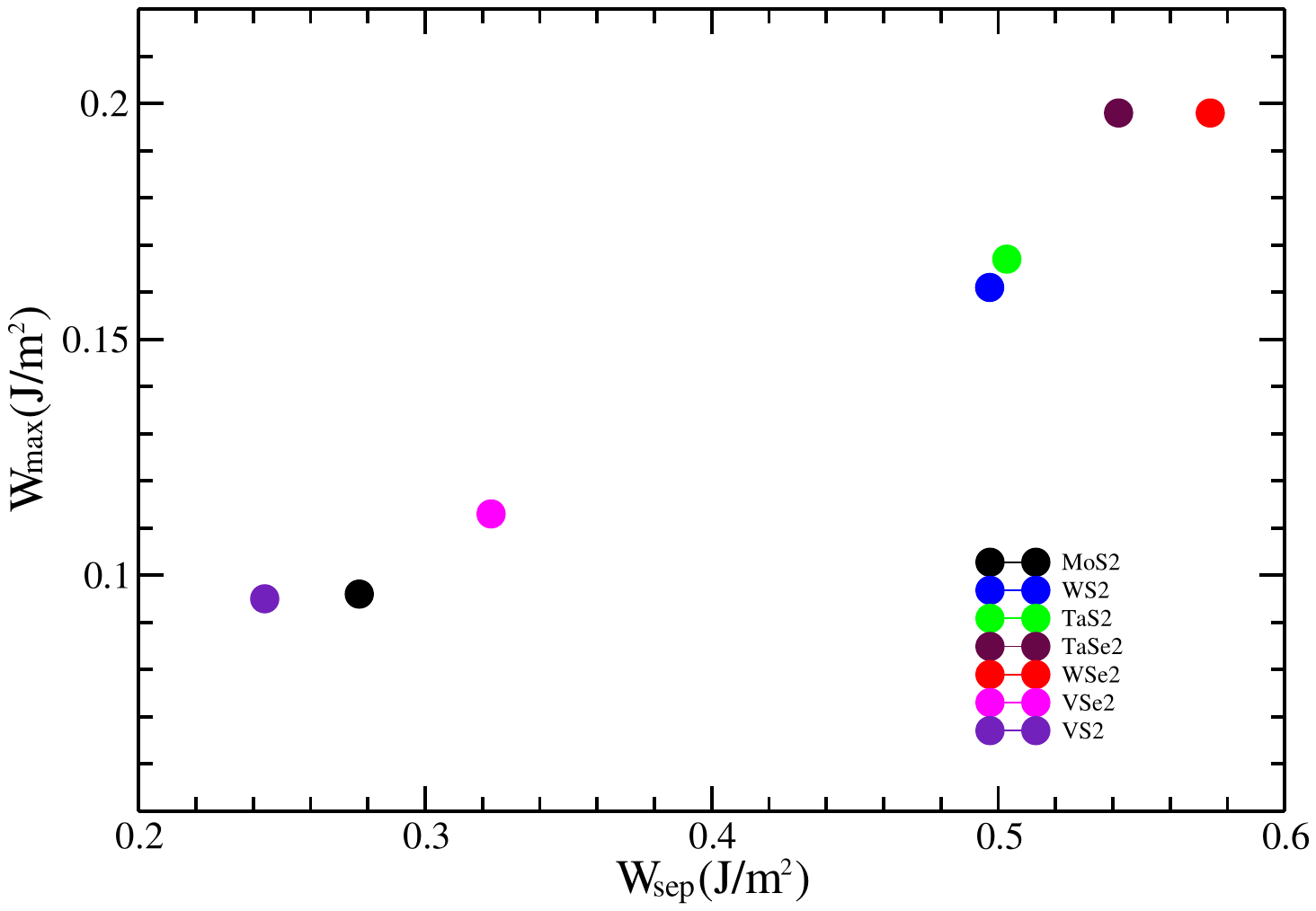}
\caption{Maximum sliding corrugation $W_{\rm max}$ versus
work of separation $W_{\rm sep}$ for selected TMD bilayers 
with the R180 structure in the {\it 2H-octahedral} configuration.}
\label{fig7}
\end{figure}

Moreover, Fig. 8 and Fig. S3 (in supplementary material) suggest 
that both $\rho_{\rm redist}$
and the maximum corrugation $W_{\rm max}$ are proportional to the lattice 
constant of the system,
by changing the chalcogen element, so that friction seems to increase as
a function of the size of the chalcogen element (from S to Se), 
in line with the results of previous DFT 
calculations,\cite{Levita,Irving} most likely due to the increased 
Pauli repulsion.
Apparently in ref.\onlinecite{Carpick} the opposite effect was found,
however it must be pointed out that the nanoscale friction behavior
reported in that study was obtained by a different approach, 
considering ``a tip sliding on a TMD''
instead of ``sliding between two TMD layers''. 
Note again the peculiar VS$_2$ and VSe$_2$ cases: $\rho_{\rm redist}$
is larger in VS$_2$ than in VSe$_2$, in spite of the fact that the
lattice constant of VS$_2$ is smaller.
The critical role of the lattice constant in determining 
the frictional behavior at the nanoscale is in line with previous
investigations.\cite{Carpick}

\begin{figure}
\centerline{
\includegraphics[width=8cm,angle=270]{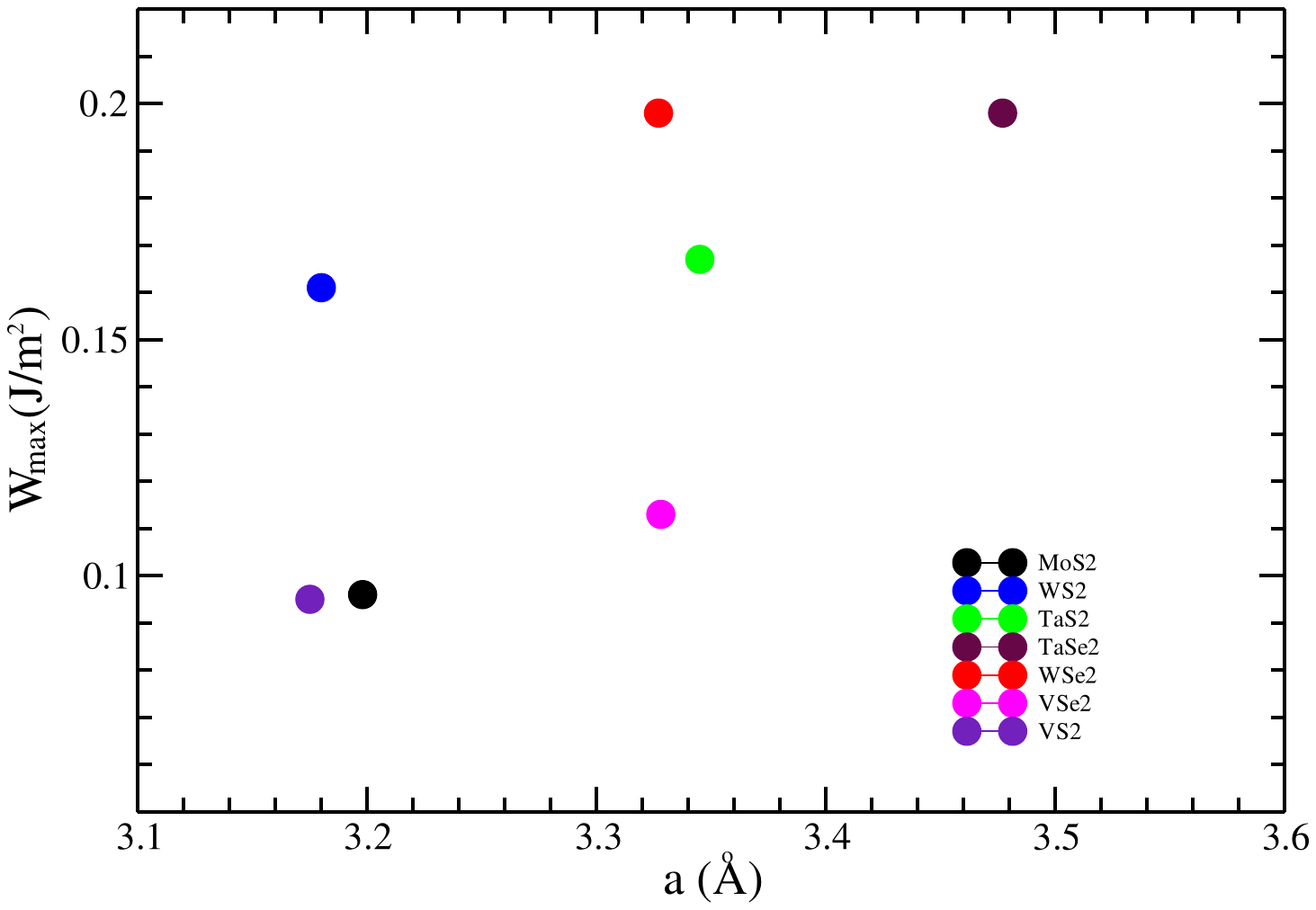}
}
\caption{Maximum sliding corrugation $W_{\rm max}$ versus
lattice constant $a$ for selected TMD bilayers 
with the R180 structure in the {\it 2H-octahedral} configuration.}
\label{fig8}
\end{figure}

We have also cosidered a pair of {\it heterostuctures}, namely two
bilayers each composed of two different TMD monolayers: MoS$_2$WS$_2$
and MoS$_2$VS$_2$.
As can be seen in Table IV and Figs. 9 and 10, MoS$_2$WS$_2$ is
characterized by $W_{\rm sep}$ and $W_{\rm max}$ values intermediate
between those of the MoS$_2$ and WS$_2$ bilayers, while instead
in MoS$_2$VS$_2$ $W_{\rm sep}$ and $W_{\rm max}$ are smaller and 
larger, respectively than those of the MoS$_2$ and VS$_2$ bilayers,
again showing a peculiar behavior of TMD systems containing the V
element.  

\begin{table}
\vfill
\eject
\caption{Calculated values for the lattice constant $a$, 
work of separation $W_{\rm sep}$, and maximum 
sliding corrugation $W_{\rm max}$ for two bilayers composed of different
monolayers ({\it heterostructures}) in {\it 2H-octahedral} configuration
with R0 orientation, compared to data relative to the 
corresponding bilayers composed of identical monolayers (homostructures).
$a$ is in \AA\ and energies in J/m$^2$.}
\begin{ruledtabular}
\begin{tabular}{|c|c|c|c|}
        system  & $a$   &  $W_{\rm sep}$ & $W_{\rm max}$ \\ 
\hline
 MoS$_2$WS$_2$ & 3.184 & 0.368 & 0.129 \\
 MoS$_2$VS$_2$ & 3.175 & 0.199 & 0.110 \\
\hline
 MoS$_2$       & 3.198 & 0.277 & 0.100 \\
 WS$_2$        & 3.180 & 0.497 & 0.167 \\
 VS$_2$        & 3.175 & 0.247 & 0.091 \\
\end{tabular}
\end{ruledtabular}
\label{table4}
\end{table}

\begin{figure}
\includegraphics[width=8cm,angle=270]{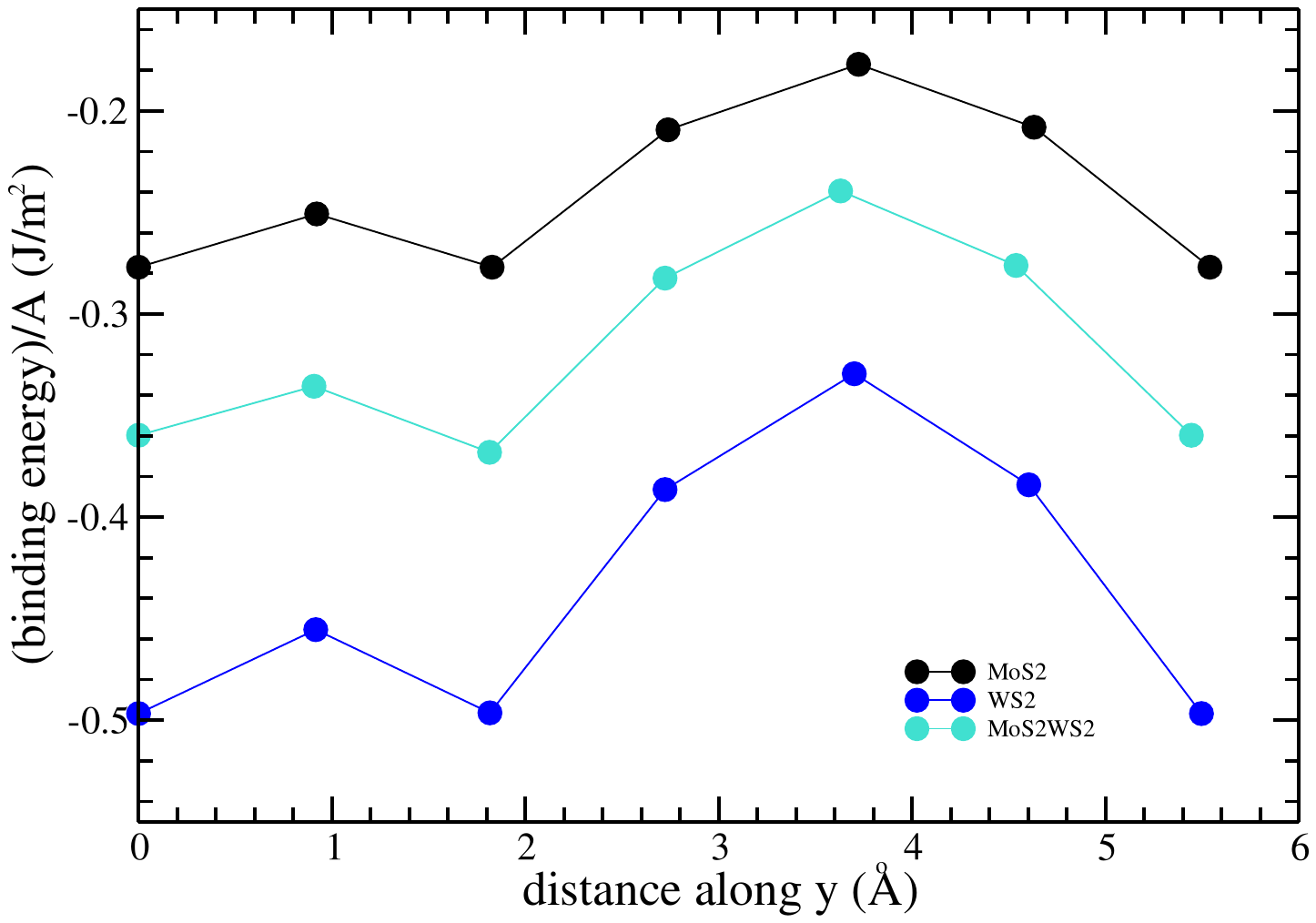}
\caption{Variation of the interlayer binding energy
per unit-cell area, $E_b/A$, in
MoS$_2$, WS$_2$ bilayers, and MoS$_2$WS$_2$ heterostructure, 
considering the R0 orientation and the {\it 2H-octahedral} configuration. 
Symbols correspond to calculated data; lines are just guides for the eyes.}
\label{fig9}
\end{figure}

\begin{figure}
\includegraphics[width=8cm,angle=270]{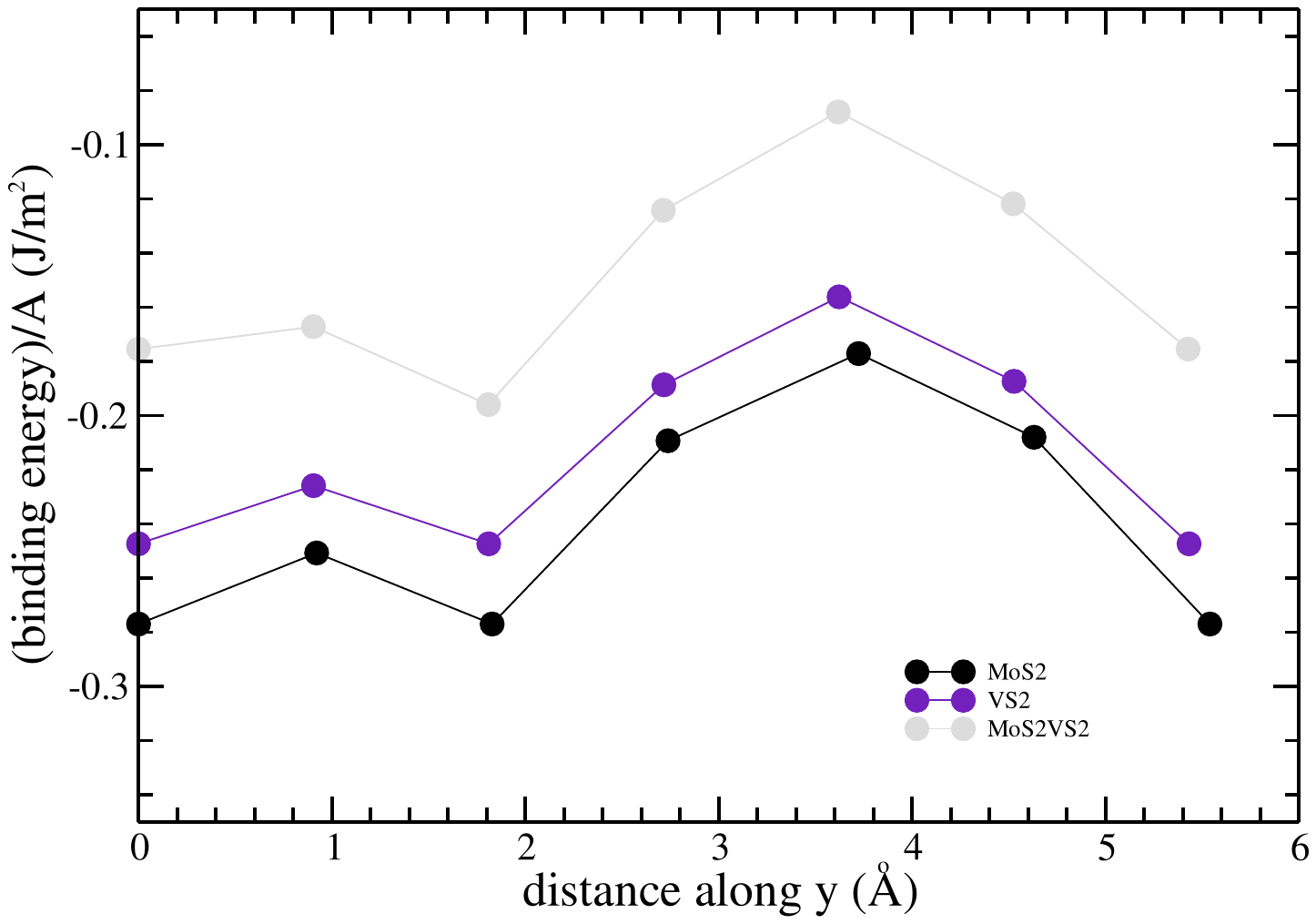}
\caption{Variation of the interlayer binding energy
per unit-cell area, $E_b/A$, in
MoS$_2$, VS$_2$ bilayers, and MoS$_2$VS$_2$ heterostructure,
 considering the R0 orientation and the {\it 2H-octahedral} configuration. 
Symbols correspond to calculated data; lines are just guides for the eyes.}
\label{fig10}
\end{figure}

\section{Conclusions}
\noindent
We have presented the results of a first principles, DFT investigation, 
of the energetic and electronic properties of several TMD bilayers
by particularly focusing on their sliding properties. 
The R180 structure appears to be preferred by most of
systems in the {\it 2H-octahedral} configuration, while the R0
structure is instead favored by most of the systems in
the {\it 1T-trigonal} configuration.
Our investigation substantiates the prevalence of van der Waals (vdW) 
interactions by quantifying their contribution to interlayer cohesion. 
We also assessed the impact of different vdW-corrected Density Functional 
Theory (DFT) functionals on our numerical outcomes.
Using MoS$_2$ as a benchmark, we observed that most other TMDs exhibit enhanced 
interlayer bonding and corrugation. Notably, TiSe$_2$
displays characteristics akin to MoS$_2$, while TiS$_2$, VS$_2$, and ZrS$_2$
demonstrate reduced bonding strength and corrugation compared to MoS$_2$,
which makes them promising for applications in fields such as nanoelectronics, 
optoelectronics, and sensing technologies. 
Interestingly, the corrugation and thus the energy barriers for 
interlayer sliding generally increases with the size of the 
chalcogen element; moreover, it is even slightly lower than that of graphene 
for selected configurations of TiS$_2$ and ZrS$_2$ bilayers. 
We also investigated correlations among various TMD bilayer properties, with 
particular emphasis on potential links between tribological and electronic 
characteristics often associated with solid interfaces, considering
adhesion energies, interfacial charge density redistributions, 
differential charge densities, and corrugation patterns. We found that,
while the corrugation is typically proportional to the adhesion energy,
the interplay between most structural, energetic, and electronic properties 
does not adhere to a single, clearly defined pattern. This complexity 
underscores the multifaceted nature of TMD bilayer systems and highlights 
the need for comprehensive studies to fully understand their behavior.
The observation of a strong positive correlation 
between $W_{\rm sep}$ and the maximum sliding corrugation $W_{\rm max}$
($W_{\rm max} \sim 0.35 W_{\rm sep}$) has important
implications for further theoretical and experimental
investigations aimed at developing a comprehensive understanding of the 
tribological properties of TMDs: the effect of
external factors, such as applied pressure, temperature, or the presence
of defects or impurities, on the $W_{\rm sep}$-$W_{\rm max}$ correlation
could be studied to understand their influence on the sliding behavior;
the correlation could be investigated under different environmental
conditions (for example in the presence humidity); experiments could be
designed to study the evolution of the $W_{\rm sep}$-$W_{\rm max}$
correlation as a function of sliding distance or cycles, which could
provide insights into the wear and degradation mechanisms of TMD
interfaces.

\section{Supplementary Material}
See the supplementary material for further figures illustrating the
connection between the intrinsic tribological properties and the electronic 
properties for selected TMD bilayers
with the R180 structure in the {\it 2H-octahedral} configuration:
the work of separation $W_{\rm sep}$ versus the redistribution of 
charge density in the interface region
$\rho_{\rm redist}$ and versus the height $h$ of the central peak of 
the planar average of the differential density;
the lattice constant $a$ versus $\rho_{\rm redist}$. 

\section{Acknowledgements}
These results are part of the ”Advancing Solid Interface and Lubricants 
by First Principles Material Design (SLIDE)” project that has received 
funding from the European Research Council (ERC) under the European 
Union’s Horizon 2020 research and innovation program (Grant agreement 
No. 865633). 


\vfill
\eject

\end{document}